\documentclass[12pt]{iopart}

\usepackage{graphicx}
\usepackage{setstack} 
\usepackage{iopams}
\usepackage{color}

\begin{document}

\title[Correlated continuous time random walks]{Correlated continuous-time random walks: combining scale-invariance with long-range memory for spatial and temporal dynamics}


\author{Johannes H. P. Schulz$^\dagger$, Aleksei V. Chechkin$^{\ddagger,\P}$,
and Ralf Metzler$^{\flat,\sharp}$}
\address{$\dagger$ Physics Department, Technical University of Munich,
D-85747 Garching, Germany\\
$\ddagger$ Akhiezer Institute for Theoretical Physics, Kharkov Institute
of Physics and Technology, Kharkov 61108, Ukraine\\
$\P$ Max-Planck Institute for the Physics of Complex Systems,
N{\"o}thnitzer Stra{\ss}e 38, 01187 Dresden, Germany\\
$\flat$ Institute for Physics \& Astronomy, University of Potsdam,
D-14476 Potsdam, Germany\\
$\sharp$ Physics Department, Tampere University of Technology, FI-33101
Tampere, Finland}

\begin{abstract}
Standard continuous time random walk (CTRW) models are renewal processes in the
sense that at each jump a new, independent pair of jump length and waiting time
are chosen. Globally, anomalous diffusion emerges through action of the generalized
central limit theorem leading to scale-free forms of the jump length or waiting
time distributions. Here we present a modified version of recently proposed
correlated CTRW processes, where we incorporate a power-law correlated noise on the level of both jump length and waiting time dynamics. We obtain a very
general stochastic model, that encompasses key features of several paradigmatic
models of anomalous diffusion: discontinuous, scale-free displacements as in
L{\'e}vy flights, scale-free waiting times as in subdiffusive CTRWs, and the
long-range temporal correlations of fractional Brownian motion (FBM). We derive
the exact solutions for the single-time probability density functions and extract
the scaling behaviours. Interestingly, we find that different combinations of
the model parameters lead to indistinguishable shapes of the emerging probability
density functions and identical scaling laws. Our model will be useful to
describe recent experimental single particle tracking data, that feature a
combination of CTRW and FBM properties.
\end{abstract}


\section{Introduction}
\label{intro}

Anomalous diffusion arises in a wide range of systems across disciplines and
is usually characterized in terms of the mean squared displacement (MSD)
\begin{equation}
\label{msd}
\left<[X(t)-X(0)]^2\right>\simeq t^{2H}
\end{equation}
of a random variable $X(t)$,
where the anomalous diffusion or Hurst exponent $H$ distinguishes subdiffusion
($0<H<\frac{1}{2}$) from superdiffusion ($H>\frac{1}{2}$) \cite{Bouchaud1990,Metzler2000}.
Normal ($H=\frac{1}{2}$) and ballistic ($H=1$) motion are contained as limiting
cases. In general, stochastic modelling is the approach of choice when the extent
and complexity of a deterministic, multidimensional system prohibits analytical,
first principles treatment. The time evolution of a small subsystem (for instance,
the dispersion of a tracer particle in aquifers, a labelled molecule in a
biological cell or the price of an individual stock on the market) is described
in terms of a stochastic process $X(t)$.
Examples for anomalous diffusion of the form (\ref{msd}) range from the motion
of charge carriers in amorphous semiconductors \cite{Scher1975} over the diffusion
of submicron tracers in living biological cells \cite{Tabei2013} or the dynamics of small particles in weakly chaotic flows \cite{Solomon1993} to the dispersion of
chemical tracers in the groundwater \cite{Scher2002} or the dynamics of stock markets \cite{bouchaud1}, just to name a few \cite{Bouchaud1990,Metzler2000,Hoefling2013,FracDyn}.

In general anomalous diffusion processes are not universal and thus their
definition through equation (\ref{msd}) is not unique. Instead, the form
(\ref{msd}) may be caused by multiple physical mechanisms, some of which are
very distinct conceptually. Several pathways to anomalous diffusion have been
discussed. Among others, these include (i) trapping mechanisms leading to long
sojourn times, (ii) long-ranged temporal correlations induced by interaction
with a complex surrounding and (iii) long-distance displacements. A prominent
approach to mathematically model these effects is via stochastic processes such
as continuous time random walks (CTRWs) \cite{Scher1975,RaWRaE}, fractional Brownian motion (FBM) \cite{Mandelbrot1968}, or L{\'e}vy flights and walks \cite{Klafter1987}.
They are paradigmatic in the sense that they are designed to tackle one specific
key property (i)-(iii). Thus CTRWs were proposed as a model for charge carrier
transport in amorphous semiconductors \cite{Scher1975}, where individual charges
reside on specific, sparsely distributed acceptor sites for long, effectively
random time spans, before hopping to a neighbour site. Moreover, crosslinked polymer filament networks as found in mammalian cells can cause similar caging effects for micron-sized objects~\cite{Wong2004}, similarly to multiscale trapping times of particles on sticky surfaces \cite{chaikin}. FBM addresses the problem
of highly correlated stock market decisions~\cite{x} and telecommunications \cite{norros}, and the associated fractional Gaussian noise fuels the diffusion of a single tracer particle in viscoelastic or crowded environments \cite{Jeon2013b,Goychuk2009,Weiss2013,Jeon2012a,Sander2012}. L{\'e}vy flights and walks \cite{Klafter1987} provide a
statistical description for the motion of tracers in weakly chaotic systems~\cite{Solomon1993} or
of the linear particle diffusion along a fast-folding polymer chain~\cite{Lomholt2005}.

However, in complex, disordered environments we would expect that more than one
of the patterns (i) to (iii) emerge, compete and collude to generate anomalous
diffusion patterns, and it remains an open challenge to identify and
differentiate them. Thus, the global properties of dispersion in amorphous media can be related to the microscale flow dynamics by adding a memory component to the standard CTRW description~\cite{Anna2013}. Modern single particle tracking techniques in experiment
and simulations indeed corroborate the co-existence of different diffusion
mechanisms~\cite{Tabei2013,Hoefling2013,Weigel2011,Akimoto2011,Szymanski2009}. For instance, for the motion of individual granules in the intracellular
fluid of living cells characteristics of CTRW-style trapping and FBM-like
anticorrelations were observed \cite{Tabei2013}. We here introduce a stochastic
process, namely the correlated CTRW (CCTRW), that merges and extends the
classical paradigmatic models of CTRW, FBM, and L{\'e}vy flights. We study in
particular two quantities, which are typically accessible experimentally: the
scaling of the particle position $X$ with time $t$, and the shape of the
probability density function (PDF) $p(x;t)$ of the particle displacement $x$
at some instant of time $t$. The result is a very flexible stochastic model,
that will be of use for the data analysis of stochastic processes in complex
systems. One immediate lesson is the interplay of the underlying stochastic
modes, the blend of which may lead to indistinguishable forms for the PDF
$p(x;t)$ for different sets of model parameters.

The paper is organized as follows. In section~\ref{model} we define the
ingredients of the CCTRW model. First, subsections~\ref{lf} and~\ref{ctrw} 
recapitulate the essential definitions and properties of L{\'e}vy flights
and CTRW-trapping theory. Second, we define a correlated version of these
models by means of stochastic integration in subsection~\ref{cctrw}. The
stationarity (closely related to the physical concept of equilibrium) of
the CCTRW is briefly discussed in section~\ref{stat}. We study extensively
the scaling behaviour and the PDF of the position coordinate for correlated
CTRWs in section~\ref{pdf}. The parallels and differences with other correlated CTRW models in the literature are outlined in section~\ref{comparison}.
A brief overview on our main results and potential
extensions and further studies of the model  are summarized in the 
conclusion, section~\ref{conclusion}.

\section{Model Definition}\label{model}

\begin{figure}
\begin{center}
\includegraphics[width=14cm]{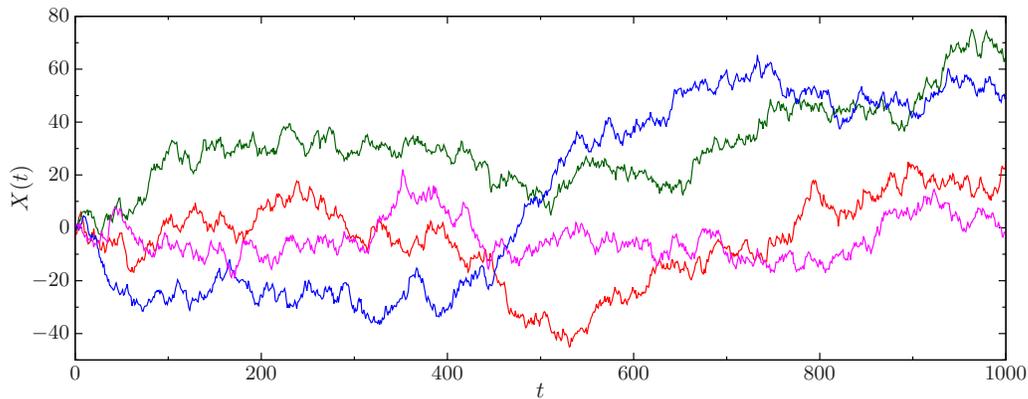}
\end{center}
\caption{Trajectories $X(t)$ of ordinary \emph{Brownian motion}. The sample 
paths are erratic but continuous and at no time a favoured direction can be determined. In the context of correlated continuous time random walks as defined 
in section~\ref{cctrw}, the parameters are: $\mu=2$, $K=1/2$, 
$\alpha=1$, and $G=1$.}
\label{fig.bm}
\end{figure}

The most commonly used theoretical model for (normal) diffusion dynamics is the
celebrated Brownian motion. In its standard form, this random process describes
the dynamics of a point-like particle as an unbiased, continuous but erratic
motion in an unbounded embedding space. In the mathematics literature, such a
process $X(t)$ with positive time $t\in\mathbb{R}_0^+$, is usually referred to
as the Wiener process, and it is uniquely defined by the following three 
properties: (i) $X(0)=0$; (ii) a sample trajectory $X(t)$ is almost surely
continuous everywhere; (iii) increments $X(t_2)-X(t_1)$ have a Gaussian
distribution with mean $0$ and variance $|t_2-t_1|$, and they are mutually 
independent for any non-overlapping time intervals. Typical sample realisations
of Brownian motion are shown in Fig.~\ref{fig.bm}. Individual trajectories are
indeed characterized by a continuous but non-smooth behaviour. There is no notable
global drift and neither a specific point in time nor some spatial region stands
out from the rest. In ensemble measurements, Brownian motion features a normal
diffusive behaviour ($H=\frac{1}{2}$). More generally, the position coordinate
scales with time as $X(t)\sim t^{1/2}$. The independence of increments is
reflected by the correlation function, $\langle X(t_1)X(t_2)\rangle=\min\{
t_1,t_2\}$ for any $t_1,t_2$. In this sense, we call $X(t)$ an uncorrelated
process.

Thus, Brownian motion is an ideal candidate to model the diffusive motion in 
an environment where the bombardment by small particles
from a surrounding heat bath induces vivid but short displacements of a relatively
inert, point-like test particle. The typical example is a micron-scale molecule
in a dilute water solution at room temperature, as discussed in Albert Einstein's
groundbreaking studies \cite{Einstein1905} and monitored in the seminal works by Jean
Perrin \cite{Perrin1909}. There, the momentum transfer from the surrounding water
molecules occurs much faster than the average large particle motion observed
under a microscope. Displacements of the test particle thus indeed appear to be
random and independent, yet small on an observational scale.

However, when we want to describe diffusive dynamics in complex environments, we 
are forced to drop several of the idealising assumptions (i) to (iii). For instance, in
chaotic systems \cite{zaslavsky} or in highly disordered optical materials
\cite{light} large scale displacements occur almost instantaneously, resulting
in highly non-continuous sample trajectories. Conversely, in disordered
environments such as the above-mentioned amorphous semiconductors or the densely
crowded intracellular fluid of biological cells, the assumption of a steady
time evolution is questionable, since charge carriers or tracer molecules can
become trapped in microenvironments or stick to reactive surfaces for long time
periods~\cite{Wong2004}. Finally, the independence of increments cannot be taken for granted when the particle motion is strongly coupled to its environment, for instance, in single file diffusion~\cite{Sander2012} or in viscoelastic media~\cite{Goychuk2009}, leading to an effective memory in the history of the motion.

In the following sections, we generalise the standard Brownian motion to a 
larger class of one-dimensional, random motions of a point particle to accommodate
the above-mentioned effects. We proceed stepwise: first, in section~\ref{lf}, we
introduce the concept of L\'evy flights (allowing discontinuities in sample
paths). In section~\ref{ctrw} we consider CTRWs (with long sojourn times).
Finally in section~\ref{cctrw}, we generalise to processes defined in terms of 
stochastic integrals (to account for FBM-style memory effects). On each level of 
generalisation, we focus our analytical discussion on the study of scaling laws 
in general, and the evolution of the PDF of the particle position $x$ with
respect to time $t$, in particular.

\subsection{L\'evy flights}\label{lf}

\begin{figure}
\begin{center}
\includegraphics[width=14cm]{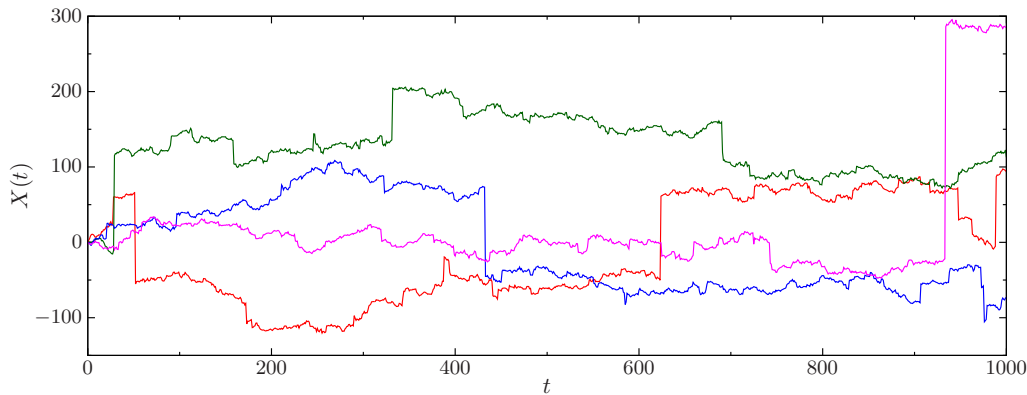}
\end{center}
\caption{Sample trajectories $X(t)$ of \emph{L\'evy flights}. The motion is 
uncorrelated and unbiased, but characterized by large-scale, discontinuous 
jumps. In the context of CCTRWs as defined in section~\ref{cctrw}, the
parameters are: $\mu=3/2$, $K=2/3$, $\alpha=1$, and $G=1$.}
\label{fig.lf}
\end{figure}

The first assumption that we drop in order to extend ordinary Brownian
motion is the continuity of sample paths. A widely used stochastic approach to
model such type of anomalous diffusion property are L{\'e}vy flights. This type
of idealized random motion is not continuous in space, but instead consists of
a series of random jumps, see Fig.~\ref{fig.lf}. The jump lengths $\delta x$
are characterized by heavy-tailed jump statistics, that is, by PDFs with
power-law tails, $\lambda(\delta x)\simeq|\delta x|^{-1-\mu}$ with $0<\mu<2$
for large distances $|\delta x|$. The key feature of such stable PDFs is the
diverging second moment, $\langle(\delta x)^2\rangle=\int_{-\infty}^\infty
(\delta x)^2\lambda(\delta x)\mathrm{d}(\delta x)=\infty$: discontinuous jumps
occur on arbitrary large spatial scales. 

To be more precise, we here connect L{\'e}vy flights with the mathematical
concept of a symmetric L{\'e}vy $\mu$-stable process, $X(t)=L_\mu(t)$. Apart
from the discontinuity of individual trajectories, also ensemble properties
differ significantly from the Brownian case: the capability of covering large
distances by single, instantaneous jumps implies a superdiffusive scaling of
the position coordinate with time, $X(t)\sim t^{1/\mu}$. In particular, the
PDF $p_\mu(x;t)$ for the position $X$ at time $t$ assumes the scaling form
\begin{equation}
p_\mu(x;t)=t^{-1/\mu}\ell_\mu\left(xt^{-1/\mu}\right).
\label{symstable}
\end{equation}
where the scaling function $\ell_\mu$ is a symmetric $\mu$-stable law. The 
latter is uniquely defined in terms of the characteristic function
\begin{equation}
\left<\exp\left\{ikL_\mu(1)\right\}\right>=\int_{-\infty}^\infty e^{ikx}\ell_
\mu(x)\mathrm{d}x=\exp\left(-|k|^\mu\right),
\label{symstable_fourier}
\end{equation}
including Gaussian statistics in the limit $\mu=2$. Indeed, in a distributional
sense, the limiting case $X(t)=L_2(t)$ is an ordinary Brownian motion. The
scaling function $\ell_\mu$ (and thus the PDF $p_\mu$) has the same heavy-tail
property as the PDF of individual jumps, $\ell_\mu(x)\simeq|x|^{-1-\mu}$ for
large $|x|$. Consequently, second (and higher) order moments diverge, $\langle
X^2(t)\rangle=\int_{-\infty}^\infty x^2p_\mu(x;t)\mathrm{d}x=\infty$.

Yet, in terms of correlations, L{\'e}vy flights are on the same level as
ordinary Brownian motion. Indeed, both processes are Markovian. The jump
lengths $\delta x$ are mutually independent and identical in a distributional
sense. Consequently, the increments of L{\'e}vy stable motion $X(t_2)-X(t_1)$,
are characterized by mutual independence and distributional equality.

\subsection{Continuous time random walks}\label{ctrw}

\begin{figure}
\begin{center}
\includegraphics[width=14cm]{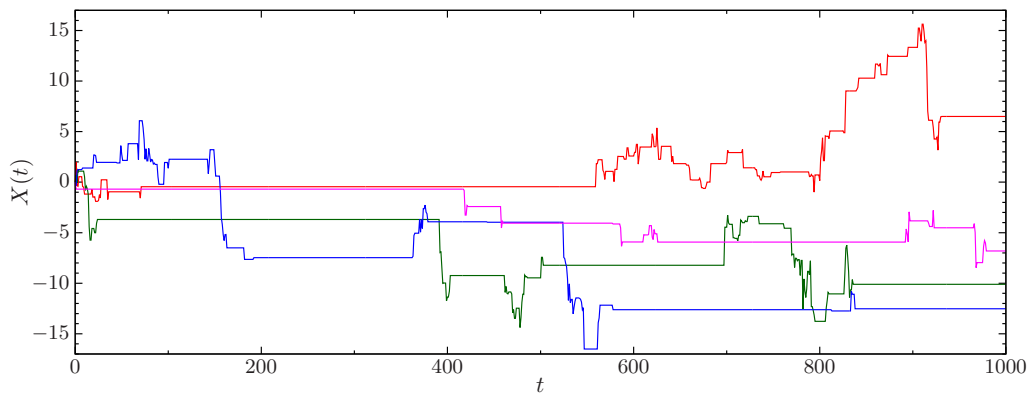}
\end{center}
\caption{Sample trajectories $X(t)$ of \emph{continuous time random walks 
(CTRWs)}. The spatially continuous motion is paused for scale-free waiting
periods. This considerably slows down the exploration of space as compared to 
ordinary Brownian motion. Waiting times are heavy-tailed (see text) and thus 
assume values on all time scales, but they are also mutually independent. In the 
context of CCTRWs as defined in section~\ref{cctrw}, the parameters are:
$\mu=2$, $H=1/2$, $\alpha=1/2$, and $G=2$.}
\label{fig.ctrw}
\end{figure}

Despite their spatial discontinuity, L{\'e}vy flights are evolving continuously
in time, in the sense that the particle remains at any specific position for only
an infinitesimal amount of time. To account for the possibility to encounter
deep traps on some random energy landscapes, a common generalized stochastic
model is used, namely, subdiffusive CTRWs. In contrast to ordinary Brownian
motion or L{\'e}vy flights, these processes include random long-time trapping
periods $\delta t$, usually referred to as waiting times. These are distributed
according to heavy-tailed waiting time statistics, $\psi(\delta t)\simeq \delta
t^{-1-\alpha}$ with $0<\alpha<1$ for large $\delta t$. In close analogy to the
effects of scale-free jump lengths for L\'evy flights, the infinite first moment
of the waiting times amounts to immobilisation periods on all time scales, see
Fig.~\ref{fig.ctrw}.

Mathematically, one can model such systems by use of the subordination method 
\cite{Fogedby1994,Meerschaert2004}. An internal time parameter $s$ is introduced, which plays the role of
the number of jumps performed along the trajectory, that is successively delayed
by trapping events. As this `internal time' $s$ increases, on the one hand, the
spatial exploration evolves according to a process $Y(s)$. The appropriate choice
for $Y(s)$ depends on the characteristics and features of the physical system we
intend to describe (external force fields, drift, friction, etc.). Typically,
$Y(s)$ is assumed to be Markovian and thus, in the above sense, is continuously
evolving in time. As a paradigmatic example, we define $Y(s)=L_\mu(s)$, i.e.,
the spatial dynamics are modelled in terms of an unbiased and unconfined L{\'e}vy
flight with stable index $0<\mu<2$, as defined in the previous section~\ref{lf}. 

On the other hand, we model the punctuated progression of real laboratory time as 
measured by the observer in terms of a separate random process $T(s)$: as the
internal time $s$ increases, consecutive waiting times accumulate to the 
laboratory time $T(s)$. In order to model waiting times which are distributed
by heavy-tailed statistics, one can simply choose $T(s)=L^+_\alpha(s)$, where
$0<\alpha<1$. The latter is a special type of L{\'e}vy flight itself, namely,
a one-sided (or totally skewed) L{\'e}vy $\alpha$-stable motion. It is a positive,
strictly increasing process and thus an appropriate representation of the random
time progression of the particle motion. Moreover, it has the typical L{\'e}vy
flight property of scale-free discontinuous, jump-like evolution. Statistically,
it is characterized by the time scaling $T(s)\sim s^{1/\alpha}$ and one-sided 
$\alpha$-stable distributions. In particular, the PDF $g_\alpha(t;s)$ for the 
laboratory time $T$ at given internal time $s$ reads
\begin{equation}
g_\alpha(t;s)=s^{-1/\alpha}\ell_\alpha^+(ts^{-1/\alpha}).
\label{osstable}
\end{equation}
Here, $\ell_\alpha^+$ is a one-sided $\alpha$-stable law. Its most natural 
representation is via its Laplace transform
\begin{equation}
\left<\exp\left\{-\theta L_\alpha^+(1)\right\}\right>=\int_0^\infty e^{-\theta
t}\ell_\alpha^+(t)\mathrm{d}t=\exp\left(-\theta^\alpha\right) .
\label{osstable_laplace}
\end{equation}
The distribution is heavy-tailed, $\ell_\alpha^+(t)\simeq t^{-1-\alpha}$, so 
that the expectation value (and higher order moments) of the laboratory time 
diverges, $\langle T(s)\rangle=\int_0^\infty tg_\alpha(t;s)\mathrm{d}t=\infty$.
Note that in the limit $\alpha\rightarrow1$ the PDF in Eq.~(\ref{osstable})
becomes a Dirac $\delta$-distribution, $g_1(t;s)=\delta(t-s)$. Thus, this
limiting case restores the equivalence of internal and laboratory time, such
that the particle motion is no longer paused for random waiting time periods.

To complete the definition of this type of CTRWs we introduce an inverse process
$S(t)$ which measures the evolution of internal time as function of laboratory
time $t$,
\begin{equation}
S(t)=\inf\{s>0:T(s)>t\},
\label{S}
\end{equation}
which is also sometimes referred to as first-hitting time or counting process. 
The particle motion as seen by the observer is now given by a process
$X(t)=Y(S(t))$, i.e. the random, unsteady progression of internal time is
modelled by $S(t)$, while independently the spatial displacements during
times of dynamic activity are governed by the process $Y(s)$. Individual paths 
of $X(t)$ most notably feature discontinuities in both their spatial and 
temporal evolution. Meanwhile, ensemble statistics combine the distributional 
properties of both independent random processes $Y(s)$ and $S(t)$. For 
instance, let $h_\alpha(s;t)$ denote the PDF for internal time $S$ at laboratory 
time $t$. Recall that $p_\mu(y;s)$ is the PDF of the position $Y$ when the 
internal time $s$ has passed. Then the PDF $p_{\mu,\alpha}(x;t)$ for the 
particle position $X$ at time $t$ can by computed as
\begin{equation}
p_{\mu,\alpha}(x;t)=\int_0^\infty p_\mu(x;s)h_\alpha(s;t)\mathrm{d}s.
\end{equation}

We conclude this section with a remark on correlations in this type of CTRW.
Both the displacement dynamics $Y(s)$ and the laboratory time evolution $T(s)$ 
with respect to internal time $s$ belong to the class of L{\'e}vy flight 
processes. As such, their respective increments are stationary and mutually 
independent for non-overlapping time intervals. In the language of individual 
jump distances $\delta x$ or waiting times $\delta t$ this means that the 
latter form sequences of mutually independent, identically distributed random 
variables. As mentioned above, this renewal property might be considered a
severe simplification when we want to model real physical systems. We will 
therefore drop this property in the following section and define a CTRW where
successive jump lengths or waiting times are correlated.

\subsection{Correlated continuous time random walks}\label{cctrw}

In standard CTRW models, individual jump lengths and waiting times, respectively,
are independent of each other. Our goal is to extend this theory to systems where
highly complex environments induce long-range correlations. In that we build on previous results \cite{Meerschaert2009,Tejedor2010} for non-renewal CTRWs in which successive waiting times and/or jump lengths are incrementally varied, such that they represent Markovian random walks in the associated spaces of waiting times and jump lengths. We discuss the parallels with this incremental-correlated CTRW in section~\ref{comparison}. In the present paper, we follow an idea proposed in~\cite{Chechkin2009} to introduce a process close in spirit to FBM and its 
heavy-tailed generalisation, the linear fractional $\mu$-stable motion
\cite{StableProc}. The basic theoretical approach is to derive a correlated process from an uncorrelated one in terms of (stable) stochastic integrals. By this method, correlations are introduced without altering the distributional properties of the process themselves.

Without going into the details of stochastic integrals \cite{StableProc} we here
provide an exemplary calculation as a motivation for our method. Consider
first a discrete time random walk in continuous space, $Y_n\in\mathbb{R}$
with $n\in\mathbb{N}$. At the $n$-th step, the random walker covers a
random jump distance $\delta x_n=Y_{n+1}-Y_n$. If we assume the $\delta x_n$
are mutually independent and identically distributed (iid), we call the random
walk $Y_n$ an uncorrelated process. For example, define $\delta x_n=\xi_n$,
where the $\xi_n$ are Gaussian iid random variables with zero mean and variance
$\sigma^2$. Then the distribution of the position variable $Y_n$ follows as
(we assume $Y_0=0$)
\begin{equation}
 Y_n = \sum_{j=1}^{n} \delta x_j = \sum_{j=1}^{n} \xi_j \overset{d}{=}
 n^{1/2} \cdot \xi_1 .
\end{equation}
Here, $\overset{d}{=}$ denotes an equality in distribution, and thus the 
position $Y_n$ after the $n$-th step also has a Gaussian distribution, scaling 
as $Y_n\sim n^{1/2}$. Consequently the ``diffusion'' law, $\langle Y_n^2\rangle
=\sigma^2 n$, is normal for this simple, uncorrelated random walk. Note that
this result is a generic one, since, by virtue of the central limit theorem,
any series of iid random displacements $\xi_n$ with zero mean and finite
variance produces asymptotically Gaussian behaviour on sufficiently large scales.

Now, how can we add correlations to this simple random walk process, without
altering its Gaussian nature? One method is by means of a linear transformation,
as proposed in reference \cite{Chechkin2009}. Let $\xi_n$ be iid Gaussian random
variables as above. We introduce a nonrandom function $M_{k}$, which we will 
refer to as correlation kernel, and define the correlated jump lengths $\delta 
\tilde{x}_n=\sum_{k=1}^{n} M_{n-k+1}\xi_k$. The latter have strong similarities
in distribution with their uncorrelated counterparts $\delta x_n$: both are
sequences of Gaussian random variables centred at zero. However, the correlated
sequence is not necessarily stationary, since $\langle\delta\tilde{x}_n^2\rangle
=\tilde{\sigma}^2_n=\sum_{k=1}^n M_k^2$. More severely, the $\delta x_n$ are by
definition mutually independent, while for the correlated sequence we have, for
any $n,m\in\mathbb{N}$,
\begin{equation}
\langle\delta\tilde{x}_n\delta\tilde{x}_{n+m}\rangle=\sum_{k=1}^{n}\sum_{l=1}^{
n+m}M_{n-k+1}M_{n+m-l+1}\langle\xi_k\xi_l\rangle=\sigma^2\sum_{k=1}^{n}M_{k}M_{
k+m}.
\label{covariance}
\end{equation}
Depending on the exact behaviour of the correlation kernel $M_k$, this
covariance function can have either a negative or positive sign, where a
positive (negative) covariance function indicates a tendency for any two
jumps to go in the same (opposite) direction. We can therefore say that the
jump lengths are either \textit{persistent\/} or \textit{antipersistent},
respectively. Only by choosing $M_k=\delta_{1k}$, $\delta_{ik}$ denoting the
Kronecker symbol, the $\delta\tilde{x}_n$ become mutually independent.

The random walk process $\widetilde{Y}_n$ associated with such correlated jump 
lengths has the following properties (again, assume $\widetilde{Y}_0=0$):
\begin{eqnarray}
\widetilde{Y}_n&=&\sum_{j=1}^n\delta\tilde{x}_j=\sum_{k=1}^n\widetilde{M}_{n-k+1}
\xi_k\overset{d}{=}\xi_1\cdot\left(\sum_{k=1}^n\widetilde{M}_{k}^2\right)^{1/2}
\label{lintrafo}\\
\widetilde{M}_k&=&\sum_{l=1}^kM_l.
\nonumber
\end{eqnarray}
Thus, the correlations indeed preserve the Gaussian nature of the process.
$\widetilde{Y}_n$ can itself be written as a linear transformation of the iid
Gaussian variables $\xi_k$ in terms of the correlation kernel $\widetilde{M}_k$.
Note that we altered the scaling behaviour with the introduction of correlations:
in contrast to the normal scaling $Y_n\sim n^{1/2}$, the scaling of the process
$\widetilde{Y}_n$ is more complex in general and depends on the exact form of
the kernel $\widetilde{M}_k$.

This method of correlating a Gaussian random walk can be readily transferred to a
time-continuous process such as the L{\'e}vy flights and CTRWs as defined in the
previous sections.\footnote{How and when a time-discrete correlated random walk
\textit{converges} to a time-continuous correlated motion is taken up in
references \cite{Meerschaert2009,Chechkin2000}.}
Let $L_\mu(s)$ be a L{\'e}vy flight with stable index $0<\mu<2$. In equation
(\ref{lintrafo}) we now substitute the discrete step number $n$ by a continuous
(internal) time variable $s$ and replace the sum over the Gaussian iid random
variables by a stochastic integration with respect to the L{\'e}vy stable noise
$\mathrm{d}L_\mu(s)$. For the correlation kernel $\widetilde{M}(s)$ we choose a
power-law, so that correlations are potentially of a long-ranged kind. That is,
we define a stochastic process $Y(s)$ in terms of a stable integral (dropping the tilde in the following):
\begin{equation}
Y(s):=(\mu K)^{1/\mu}\int_0^s(s-s')^{K-1/\mu}\mathrm{d}L_\mu(s').
\label{Y}
\end{equation}
Here $0<\mu<2$, and we will refer to $K>0$ as the Hurst exponent. With this
definition of the process $Y$ we stay in the domain of $\mu$-stable processes.
In particular, at given time $s$, its probability density function $p_{\mu,K}
(y;s)$ is of the stable form (\ref{symstable}), albeit with an altered time
scaling $Y(s)\sim s^K$,
\begin{equation}
p_{\mu,K}(y;s)=s^{-K}\ell_\mu(ys^{-K}).
\end{equation}
The scaling prefactor $(\mu K)^{1/\mu}$ in Eq.~(\ref{Y}) makes sure that the scaling function $\ell_\mu$ is again exactly represented by the characteristic function in Eq.~(\ref{symstable_fourier}).

What really sets the process $Y(s)$ apart from the ordinary L{\'e}vy motion $L_\mu(s)$ is the stochastic dependence of increments. We may also say the noise related to $Y(s)$ is strongly correlated, or coloured. However, to assess the nature of interdependence here, we cannot use the covariance function like in Eq.~(\ref{covariance}). While the latter is a meaningful and precise measure of dependence for Gaussian processes, $\mu=2$, it is ill-defined for stable processes $\mu<2$. In reference~\cite{StableProc}, several alternative concepts to deal with the stable cases are introduced and discussed, such as covariation or codifference functions. In short, applying these analytical tools to the correlated process $Y(s)$, we find positive, long-range dependence when $K>1/\mu$, and negative, short-range dependence when $K<1/\mu$. (Compare this to the analogous discussion on linear fractional stable motion in~\cite{StableProc}. An extensive discussion of the notion of long-range dependence can be found in~\cite{Samorodnitsky2006}).

We can also supplement these considerations by spectral analysis arguments, compare also Ref.~\cite{Chechkin2000,sushin}. Consider a sample path of a L\'evy flight $L_\mu(s)$ and denote its Fourier transform by $\widehat{L}_\mu(\omega)$. Now since the stable stochastic integral~(\ref{Y}) is of a convolution form, there is a simple relation in Fourier space between the correlated noise $\mathrm{d}Y(s)$ and the L\'evy stable noise $\mathrm{d}L_\mu(s)$, namely $\mathrm{d}\widehat{Y}(\omega)\propto\mathrm{d}\widehat{L}_\mu(\omega)/(-i\omega)^{K-1/\mu}$. When comparing the two noise types in the case $K>1/\mu$ we thus find that the correlation kernel in the stable integral~(\ref{Y}) emphasises the low frequency components of the correlated noise. In a sample path of $Y(s)$, we may conceive this as a comparatively steady motion, even in the form of long-term periodic cycles. Conversely, when $K<1/\mu$, high frequencies are amplified. A sample path $Y(s)$ is then fluctuating violently as compared to an ordinary L\'evy flight.

Hence, both the analysis in terms of covariation/codifference functions and the spectral analysis support the idea of an either \emph{persistent} or  \emph{antipersistent} motion $Y(s)$. (Throughout the rest of this work, such statements are equivalent to saying that the respective increment/noise process is persistent or antipersistent.) For $K>1/\mu$, persistence effects long cycles of seemingly steady, biased motion. If $K<1/\mu$, antipersistent motion is observed as being wildly fluctuating, since strong, short range, negative memory leads to a quick succession of directional turns. The special case $K=1/\mu$ recovers ordinary L{\'e}vy flights with mutually independent jump lengths.

The last step in the definition of our CCTRW model is the introduction of
correlations of waiting times. Analogously to the above, we define
\begin{equation}
T(s):=(\alpha G)^{1/\alpha}\int_0^s(s-s')^{G-1/\alpha}\mathrm{d}L^+
_\alpha(s'),
\label{T}
\end{equation}
in terms of a stable integral with respect to one-sided L{\'e}vy $\alpha$-stable 
noise $\mathrm{d}L^+_\alpha(s)$. Here, $0<\alpha<1$ and $G\geq1/\alpha$. The
corresponding PDF $g_{\alpha,G}(t;s)$ at given internal time $s$ in this case
reads
\begin{equation}
g_{\alpha,G}(t;s)=s^{-G}\ell^+_\alpha(ts^{-G}),
\end{equation}
where the basic shape is still provided by a one-sided $\alpha$-stable law $\ell_
\alpha^+$ as defined in equation (\ref{osstable_laplace}). The scaling with internal
time $s$ in this case reads $T(s)\sim s^G$. While $T(s)$ is still an
$\alpha$-stable motion, waiting times are no longer independent. Note that for
$T(s)$ to be an increasing process, we need to require that $G\geq1/\alpha$.
Thus, correlations in waiting times are necessarily of the persistent type,
and have a tendency to increase with $s$. The only exception to this rule is
$G=1/\alpha$, a parameter setting which brings us back to heavy-tailed but
uncorrelated waiting times.

\begin{figure}
\begin{center}
\includegraphics[width=14cm]{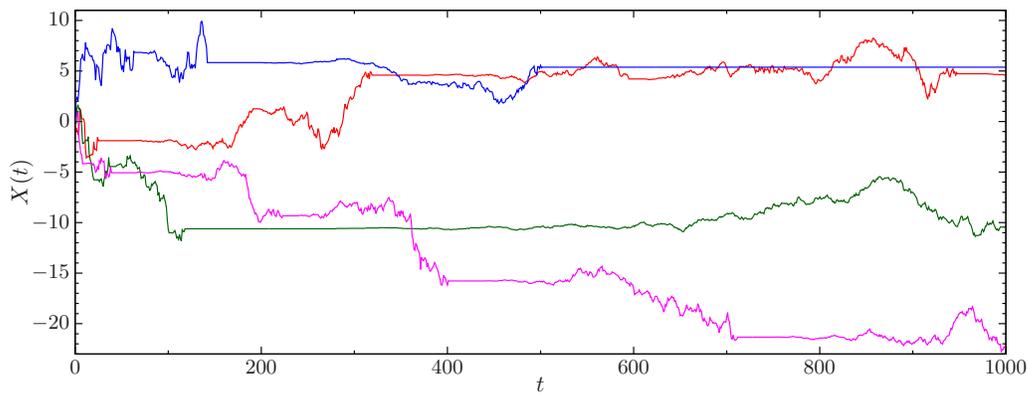}
\end{center}
\caption{Sample trajectories $X(t)$ of \emph{correlated CTRWs}. The spatially 
continuous motion is paused for large-scale waiting periods, which appear on 
all time scales. Waiting times are \emph{not\/} independent but persistent here: 
long rests are directly followed by periods of reduced dynamic activity, but 
then slowly turn into vivid almost Brownian-like motion. Note however that also 
spatial displacements are persistent. The parameters are chosen such that 
the resulting scaling with time and the shape of the PDF are the same as for the
uncorrelated CTRWs in Fig.~\ref{fig.ctrw} (as explained in section~\ref{pdf}).
In the context of CCTRWs as defined in this section, the parameters are:
$\mu=2$, $K=1.1/2$, $\alpha=1/2$, and $G=2.2$.}
\label{fig.cctrw}
\end{figure}

In complete analogy to the uncorrelated case we now introduce the inverse
process $S(t)$ according to equation (\ref{S}) and combine it with a correlated
stable motion, $X(t)=Y(S(t))$. The PDF for the particle position $X$ at real
time $t$ is then given by
\begin{equation}
p_{\mu,\alpha,K,G}(x;t)=\int_0^\infty p_{\mu,K}(x;s)h_{\alpha,G}(s;t)\mathrm{d}s,
\end{equation}
where $h_{\alpha,G}(s;t)$ denotes the PDF of internal time $S$ at real time 
$t$. We will extensively discuss this PDF in section~\ref{pdf}. To study this
process on a trajectory basis, see Fig.~\ref{fig.cctrw}.

This completes the definition of the CCTRW model that we discuss in the present
paper. A discontinuous progression of spatial displacements and laboratory time
is modelled in terms of the stable noises $\mathrm{d}L_\mu(s)$ and $\mathrm{d}L
^+_\alpha(s)$. Correlations are separately introduced by power-law correlation
kernels to both the spatial dynamics $Y(s)$ and the time evolution $T(s)$. The
full model is defined in terms of four parameters: $0<\mu<2$ and $0<\alpha<1$
determine the respective distributional properties of individual jump lengths
$\delta x$ and waiting times $\delta t$. In particular, they define the heavy
tails $\lambda(\delta x)\simeq|\delta x|^{-1-\mu}$ and $\psi(\delta t)\simeq
\delta t^{-1-\alpha}$. The special cases of continuous spatial and/or temporal
evolution are included in the correlated CTRW model on a distribution level as
the limits $\mu\rightarrow2$ and $\alpha\rightarrow1$. The parameters $K>0$ and
$G\geq1/\alpha$ directly measure the scaling exponents with respect to internal
time, $Y(s)\sim s^K$ and $T(s)\sim s^G$. Finally, the nature of the correlations
can be assessed by comparing respective parameter pairs: jump distances (waiting
times) are persistent if $K>1/\mu$ ($G>1/\alpha$), uncorrelated if $K=1/\mu$
($G=1/\alpha$), or antipersistent if $K<1/\mu$ (impossible for waiting times). 

\section{Stationarity}\label{stat}

We defined the stable processes $Y(s)$ and $T(s)$ directly in terms of their
distributional, scaling and correlation properties, as characterized through
the parameters $\mu$, $\alpha$, $K$, and $G$, respectively. We will now further
study this quite large class of processes through their stationarity properties.
For this, we apply the preliminary definition of the $n$-th order increments of
a stochastic process $Y(s)$,
\begin{eqnarray}
\nonumber
\Delta^{(1)}Y(s;\tau)&=&Y(s+\tau)-Y(s)\\
\nonumber
\Delta^{(2)}Y(s;\tau_1,\tau_2)&=&\Delta^{(1)}Y(s+\tau_2;\tau_1)-\Delta^{(1)}
Y(s;\tau_1)\\
\nonumber
&\vdots&\\
\nonumber
\Delta^{(n)}Y(s;\tau_1,\ldots,\tau_n)&=&\Delta^{(n-1)}Y(s+\tau_n;\tau_1,\ldots,
\tau_{n-1})\\
&&-\Delta^{(n-1)}Y(s;\tau_1,\ldots,\tau_{n-1}) .
\label{Y_nincrements}
\end{eqnarray}
Thus, $\Delta^{(1)}Y$ is the usual process increment while $\Delta^{(2)}Y$ is an
increment of increments, \emph{etc}. If $Y(s)$ is meant to stand for the position
of a particle at time $s$, then the ratio $\Delta^{(1)} Y(s;\tau)/\tau$ can be
viewed as the average velocity (bearing in mind that the one-time velocity, i.e.,
the limit $\tau\rightarrow0$, in general does not exist for the processes
discussed here). Likewise, $\Delta^{(2)}Y(s;\tau_1,\tau_2)/(\tau_1\tau_2)$
corresponds to the intuitive notion of an acceleration, and higher order
increments represent higher levels of temporal evolution.

We now say the $n$-th order increments of $Y(s)$ are asymptotically stationary in
distribution (ASD), if the random variable $\Delta^{(n)}Y(s;\tau_1,\ldots,\tau
_n)$ has a nontrivial limiting distribution for large times, $s\rightarrow
\infty$. In the following we will determine such degrees of stationarity for the
stable processes $Y(s)$ and $T(s)$ as defined in the previous section. Note that
this classification is not a purely academic one. For the application and
interpretation of a stochastic process as a real world model system, stationarity
properties are highly relevant. Let, for instance, $Y(s)$ model an animal
foraging process. Then stationarity of first order increments is an indication
for a time-independent search strategy: the distance $\Delta^{(1)}Y$ travelled
during, say, $\tau=1\mbox{ day}$ is statistically indistinguishable from one day
to the next. Conversely, nonstationary statistics of travel distances can be a
signature for an adaptive search strategy, an aging animal, or changes in the
environment. In this case, we could further ask whether or not such internal or
external variations are stationary. This relates to second order increments. On
smaller scales, $Y(s)$ could be a model for particle diffusion in a heat bath.
There, nonstationarity of first order increments is the fingerprint either of an
inhomogeneous environment (i.e., the particle displacement statistics changes
as the particle explores various spatial regions) or a non-equilibrated
environment (i.e., the noise imposed by interaction with the surrounding heat
bath is itself nonstationary). Then, analysis of second and higher order
increments yields information on the precise nature of the spatial or temporal
variations in the surroundings.

The displacement process $Y(s)$ as defined through the stable integral (\ref{Y})
is a nonstationary process, as indicated by the time scaling $Y(s)\sim s^K$.
As the particle explores its surrounding space, the probability to find it in any region of fixed size around the origin of motion is decaying with time. Now, the integral representation of first order increments reads
\begin{eqnarray}
\nonumber
\Delta^{(1)} Y(s;\tau)&=&(\mu K)^{1/\mu}\left\{\int_0^s\left[(s+\tau
-s')^{K-1/\mu} -(s-s')^{K-1/\mu}\right]\mathrm{d}L_\mu(s')\right.\\
&&\quad\left.+\int_s^{s+\tau}(s+\tau-s')^{K-1/\mu}\mathrm{d}L_\mu(s')\right\},
\end{eqnarray}
so that its distribution is given in terms of the characteristic function
\begin{eqnarray}
\nonumber
&&\langle\exp(ik\Delta^{(1)} Y(s;\tau))\rangle=\\
\nonumber
&&\hspace*{0.8cm}
=\exp\left[-\mu K|k|^\mu\int_0^s\left|(s+\tau-s')^{K-1/\mu}
-(s-s')^{K-1/\mu}\right|^\mu\mathrm{d}s'\right.\\
\nonumber
&&\hspace*{2.2cm}
\left.-\mu K|k|^\mu\int_s^{s+\tau}\left|(s+\tau-s')^{K-1/
\mu}\right|^\mu\mathrm{d}s'\right]\\
&&\hspace*{0.8cm}
=\exp\left[-|k|^\mu I^{(1)}_{\mu,K}(s)-|k\tau^K|
^\mu\right],
\label{Y_Y_1increments}
\end{eqnarray}
where we used the abbreviation
\begin{equation}
I^{(1)}_{\mu,K}(s)=\mu K\int_0^{s}\left|(s'+\tau)^{K-1/\mu}-(s')^{K-1/\mu}\right|
^\mu\mathrm{d}s'.
\end{equation}
Nonstationarity is indicated by the explicit $s$-dependence of the integral
$I^{(1)}_{\mu,K}$. The latter vanishes identically if $K=1/\mu$. This is natural,
since these cases are the symmetric L{\'e}vy stable motions, $Y(s)=L_\mu(s)$,
which have stationary increments by definition. Conversely, for any $K\neq1/
\mu$, the integral differs from zero, so in general the first order increments
of the stable motion $Y(s)$ are nonstationary. However, they can still be
asymptotically stationary, depending on the parameters. The expression in the
integral $I^{(1)}_{\mu,K}$ behaves, for large $s'$, like $(s')^{\mu K-\mu-1}$. The
asymptotics at large times $s\gg\tau$ are therefore given through
\begin{equation}
I^{(1)}_{\mu,K}(s)\simeq\left\{\begin{array}{ll}\mbox{const},&\mbox{for } 0<K<1,\\
\log(s),&\mbox{for } K=1,\\
\tau^\mu s^{\mu(K-1)},&\mbox{for } K>1.
\end{array}\right.
\end{equation}
First order increments are hence ASD whenever $0<K<1$, while spreading
indefinitely when $K\geq1$. We can readily extend the procedure to the study of increments of arbitrary order, see~\ref{app:stationarity}. In general, we have to distinguish two classes of parameter settings. 

If we can find a nonnegative integer $m$ such that $K=1/\mu+m$, then all increments of order $n>m$ are stationary in distribution, and lower order increments, $n\leq m$ on average broaden. This includes the L\'evy stable motions, $m=0$, $K=1/\mu$, with stationary increments of all orders. To understand this, recall that correlated and L\'evy stable noises are related in Fourier space through $\mathrm{d}\widehat{Y}(\omega)=\mathrm{d}\widehat{L}_\mu(\omega)/(-i\omega)^{K-1/\mu}$. Now for $K=1/\mu+m$, this suggests we can interpret $Y(s)$ as an $m$-fold repeated integration of a L\'evy stable noise. In other words, for $m=0$, $Y(s)$ is a L\'evy flight, so increments are stationary; for $m=1$, the noise generating $Y(s)$ is already a L\'evy flight, therefore only second and higher order increments of $Y(s)$ are stationary; for $m=2$, the noise generating the noise of $Y(s)$ is a  L\'evy flight, so we find stationary third order increments; \textit{etc}.

The opposite case is $K\neq1/\mu+m$ for all nonnegative integers $m$. Interestingly, here the result is $\mu$-independent: all increments of order $n>K$ are ASD, while lower order increments, $n\leq K$, are spreading indefinitely. An extensive and mathematically rigorous treatment of stochastic processes with stationary $n$th order increments can be found in~\cite{Yaglom1955}.

The one-sided $\alpha$-stable process (\ref{T}), which describes the evolution
of laboratory time with respect to the internal time has completely analogous
properties. Increments of any order are stationary if $G=1/\alpha$, since
then $T(s)=L^+_\alpha(s)$ is a one-sided L\'evy stable motion. If there is a nonnegative integer $m$ such that $G=1/\alpha+m$, then only increments of order $n>m$ are stationary in distribution. If there is no such $m$,  increments of orders $n>G$ are ASD. Lower order increments are nonstationary at all times. Note, however, that for the waiting time process we are forced to require $G\geq1/\alpha$ for the following reason. Writing out the integral representation for the first order increments,
\begin{eqnarray}
\nonumber
\Delta^{(1)}T(s;\tau)&=&(\alpha G)^{1/\alpha}\left\{\int_0^s\left[
(s+\tau-s')^{G-1/\alpha}-(s-s')^{G-1/\alpha}\right]\mathrm{d}L^+_\alpha(s')
\right.\\
&&\quad\left.+\int_s^{s+\tau}(s+\tau-s')^{G-1/\alpha}\mathrm{d}L^+_\alpha(s')
\right\},
\end{eqnarray}
we see that the first integral could potentially give a negative contribution
when $G<1/\alpha$. This is clearly unacceptable in terms of causality: negative
increments in laboratory time $T(s)$ would correspond to waiting times finishing
earlier than they began. We therefore consider only $G\geq1/\alpha$, which has
two implications. On the one hand, as mentioned above, correlated motions are
necessarily persistent. On the other hand, first order increments---reflecting
waiting time statistics---are nonstationary. More precisely, they are, in a
statistical sense, increasing beyond all bonds, as their (one-sided!)
distribution continuously broadens with internal time $s$.

We conclude this section with a general remark on stationarity in CTRW
models. The inverse process $S(t)$ measuring the internal time at fixed
laboratory time $t$, Eq.~(\ref{S}), is a highly nonstationary process, as are
all of its increments. This holds even when $G=1/\alpha$, i.e. when waiting
times are not correlated. This phenomenon has been discussed extensively
in the CTRW literature, where it is commonly referred to as aging~\cite{Barkai2003,Schulz2013,Burov2010}
and is closely related to other peculiar effects such as weak ergodicity
breaking~\cite{Burov2010,He2008}. The deeper reason behind this nonstationarity are scale-free characteristics of waiting times. In the context of diffusion dynamics, for instance, this absence of a typical time scale is motivated by an immense heterogeneity of the environment. In effect, the particle encounters an
indefinitely broad range of waiting times and falls into deeper and deeper
traps while exploring the environment. Thus, CTRW models are by definition
highly nonstationary stochastic processes, and it is indeed natural to
extend the common model candidates ($K=1/\mu$ for uncorrelated, stationary
jump distances and $G=1/\alpha$ for uncorrelated, stationary waiting times)
to the larger class of stable, but correlated and potentially nonstationary
motions considered here.

\section{Time scaling analysis and probability density function}\label{pdf}

For ordinary L{\'e}vy flights or CTRWs, the tail parameters $\mu$ and $\alpha$
determine both the distributional and the scaling properties of the process.
The present correlated model is slightly more complex in this respect. While
the shape of the PDF depends on all four parameters, only the Hurst parameters
$K$ and $G$ determine the time scaling. To see this, recall that for the L{\'e}vy
stable motions we have the characteristic scalings $L_\mu(s)\sim s^{1/\mu}$
and $L^+_\alpha(s)\sim s^{1/\alpha}$. From Eqs.~(\ref{Y}) and~(\ref{T})
it follows that $Y(s)\sim s^{K}$ and $T(s)\sim s^{G}$. Consequently, the
internal time scales as $S(t)\sim t^{1/G}$, and for the correlated motion we get
\begin{equation}
X(t)=Y(S(t))\sim t^{H},\quad\mbox{where }H=K/G.
\label{scaling}
\end{equation}
We therefore call $H$ the scaling or Hurst exponent of the correlated motion
$X(t)$. Interestingly, from the point of view of time scaling, persistence in
waiting times competes with persistence in jump distances, and the process can
turn out to be either sub- ($H<1/2$), or superdiffusive ($H>1/2$), or exhibit a
normal diffusive scaling ($H=1/2$). Conversely, measuring the Hurst exponent $H$
alone does not reveal specific information on the time scaling of correlated
waiting times ($G$) and correlated jumps ($K$), but only on their ratio.

This ambiguity actually goes beyond a simple time scaling analysis and extends
to the analysis of the PDF, as we show now. Let $h_{\alpha,G}(s;t)$ denote the
probability density for the internal time $S$ at given laboratory time $t$.
Recall that $T(s)\sim s^G$ is a monotonically increasing process. This implies
\cite{Meerschaert2004} $S(t)\stackrel{d}{=}(t/T(1))^{1/G}$ for any fixed laboratory
time $t$. Therefore,
\begin{equation}
h_{\alpha,G}(s;t)=Gts^{-G-1}\ell^+_\alpha(ts^{-G}).
\label{invstable}
\end{equation}
We can now combine Eqs.~(\ref{Y}), (\ref{S}) and (\ref{invstable}) to write the 
PDF $p_{\mu,\alpha,K,G}(x;t)$ for the correlated CTRW $X(t)=Y(S(t))$ at time $t$
in terms of stable densities,
\begin{eqnarray}
\nonumber
p_{\mu,\alpha,K,G}(x;t)&=&\int_0^\infty p_{\mu,K}(x;s)h_{\alpha,G}(s;t)ds\\
\nonumber
&=&\int_0^\infty\frac{1}{s^{K}}\ell_\mu\left(\frac{x}{s^{K}}\right) 
\frac{Gt}{s^{G+1}}\ell^+_\alpha\left(\frac{t}{s^{G}}\right)ds\\
\nonumber
&=&\frac{1}{t^H}\int_0^\infty\ell_\mu\left(\frac{x}{(st)^{H}}\right)\frac{1}{s
^{H+2}}\ell^+_\alpha\left(\frac{1}{s}\right)ds\\
&=:&\frac{1}{t^H}q_{\mu,\alpha,H}\left(\frac{x}{t^{H}}\right)
\label{pdfsubord}
\end{eqnarray}
This representation demonstrates that the qualitative shape of the PDF can be
classified in terms of only three parameters: the tail parameters $\mu$ and
$\alpha$ and the scaling exponent $H = K/G$. This means that two processes may
seemingly be the same when only studying their PDF and time scaling behaviour,
although they are inherently different with respect to their correlations.
Apparently, persistence in jump distances can balance persistence in waiting
times, similar to the previously observed twin paradox \cite{twin}. Consider, for instance, the stochastic process $X(t)$ defined by $\mu=2$,
$K=1/2$, $\alpha=1/2$ and $G=2$. This special case has been studied extensively
in the literature, as it represents the simplest type of a CTRW process and is
bare of correlations both in jump distances and waiting times. For comparison,
now define $X'(t)$ by choosing  $\mu'=2$, $K'=1.1/2$, $\alpha'=1/2$ and $G'=2.2$.
Obviously, $X'(t)$ is different from the ordinary CTRW $X(t)$, since both its
jump distances and its waiting times are persistent. This is
clearly visible when studying a few sample trajectories, as provided in
Figs.~\ref{fig.ctrw} and \ref{fig.cctrw} \footnote{Methods to estimate such parameters from empirical CCTRW trajectory data are discussed in~\cite{Teuerle2013}. }. However, on the level of time scaling
analysis, Eq.~(\ref{scaling}), and PDF, Eq.~(\ref{pdfsubord}), the random
motions are indistinguishable, since $H=H'$.

To analyse the PDF $p(x;t)$ analytically (we drop subscript parameters from here
on), it is natural to first study equation (\ref{pdfsubord}) in Fourier-Laplace
domain. Making direct use of equations (\ref{symstable}) and (\ref{osstable}),
we find
\begin{eqnarray}
\nonumber
p(k;u)&\equiv&\int_{-\infty}^\infty\int_0^\infty e^{ikx-ut}p(x;t)dtdx\\
&=&\int_0^\infty\exp\left(-|k|^\mu s^{H\mu/\alpha}\right)u^{\alpha-1}\exp\left(
-u^\alpha s\right)ds .
\label{pdffoulap1}
\end{eqnarray}
We now interpret the integral as a Laplace transform with respect to internal
time $s$, while expressing the exponential in terms of a Fox $H$-function~\cite{FoxH},
\begin{equation}
\exp(-z)=H^{1,0}_{0,1}\left[z\left|\begin{array}{l}\rule[0.12cm]{0.8cm}{0.01cm}
\\(0,1)\end{array}\right.\right].
\end{equation}
After some straightforward manipulations of the $H$-function~\cite{FoxH} we
arrive at the following representation in Fourier-Laplace space,
\begin{equation}
\label{pdffoulap2}
p(k;u)=\frac{\alpha}{u\mu H}H^{1,1}_{1,2}\left[\frac{u^{\alpha}}{|k|^{\alpha/H}}
\left|\begin{array}{l}(1,\alpha/(\mu H))\\(1,1)\end{array}\right.\right].
\end{equation}
Inverting to laboratory time $t$ and real space $x$, we find~\cite{FoxH}
\begin{equation}
p(x;t)=\frac{t^{-H}}{2\mu\sqrt{\pi}}H^{2,1}_{2,3}\left[\frac{|x|}{2t^H}\left|
\begin{array}{l}(1-1/\mu,1/\mu);(1-H,H)\\(0,1/2),(1-H/\alpha,H/\alpha);(1/2,
1/2)\end{array}\right.\right] .
\label{pdffoxH}
\end{equation}
Since for $H$-functions, series representations for small and large arguments
are known, we can now analyse in detail the behaviour around the origin and in
the tails. Series expansions can in principle be evaluated up to any order, see~\cite{FoxH,Kilba1999}. Here, we  discuss the leading order contributions to the PDF, or equivalently, to the scaling function $q(z)=p(z;1)$.

\begin{figure}
\begin{center}
\includegraphics[]{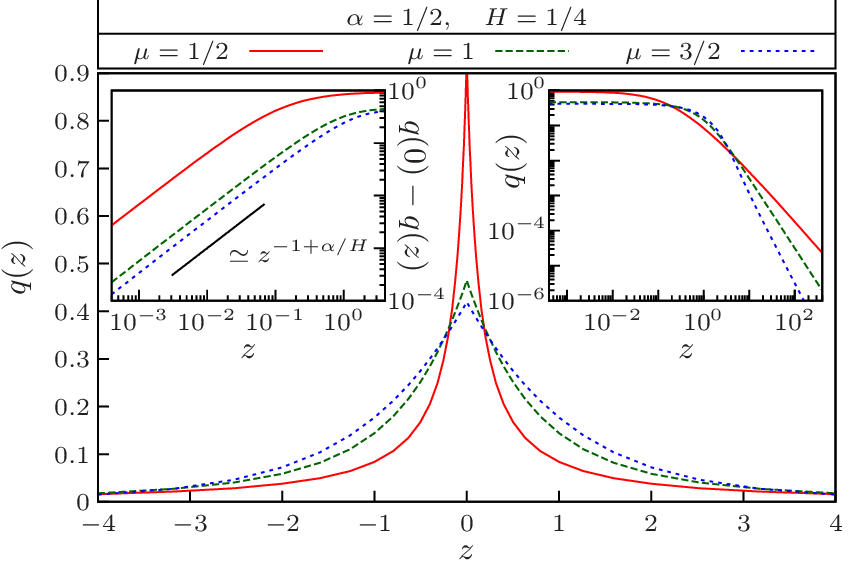}
\includegraphics[]{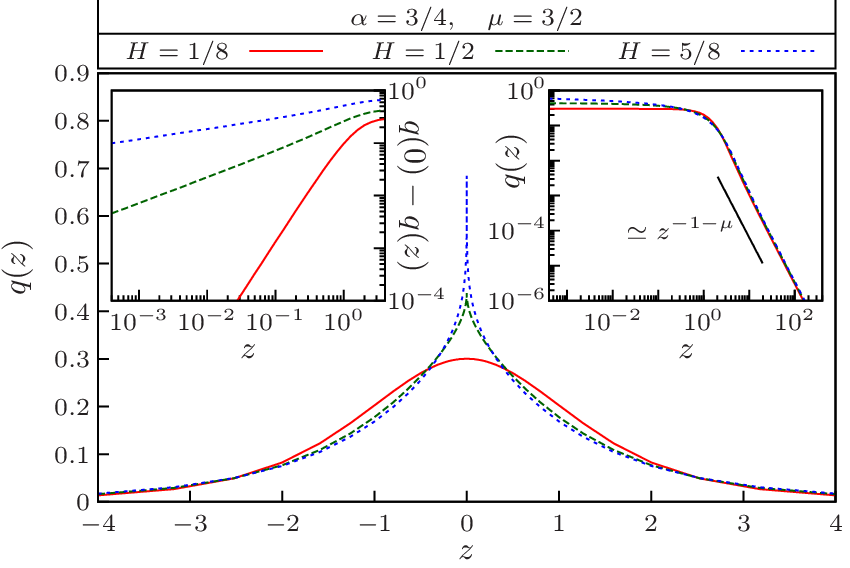}
\includegraphics[]{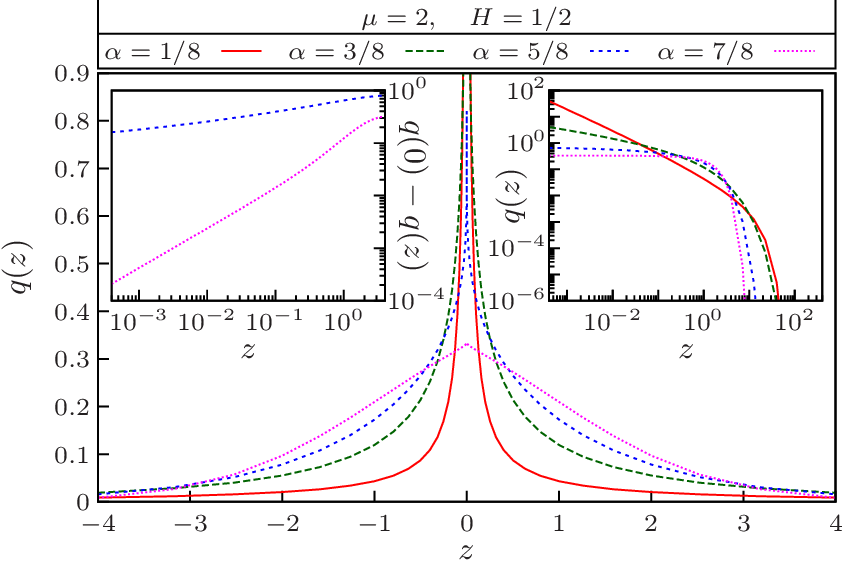}
\end{center}
\caption{Scaling function $q(z)$ for the propagator $p(x;t)=t^{-H}q(xt^{-H})$,
numerically evaluated through equation (\ref{pdfsubord}). The insets detail the
behaviour around the origin (\textit{left}) and in the tails (\textit{right}).
\textit{Top:} While the exponent of power-law tails varies with $\mu$, the
qualitative behaviour at the origin is universally given by $q(0)-q(z)\simeq
|z|^{-1-\alpha/H}$. \textit{Centre:} Conversely, a fixed stable exponent $\mu<2$
defines the tail properties, $q(z)\simeq|z|^{-1-\mu}$. By variation of the ratio
$\alpha/H$, the shape of the maximum turns from a distinct cusp to a smooth
Gaussian-like bell. \textit{Bottom:} With $\mu=2$, the tails are stretched
exponentials. When $\alpha<H$, the scaling function diverges at the origin,
$q(0)=\infty$. With $H=1/2$, an analysis of the mean squared displacement
universally indicates normal diffusion, since $X^2\sim t$.}
\label{fig.pdf}
\end{figure}

In the vicinity of the starting position, $z\approx0$, we find that the
qualitative shape depends highly on the ratio $\alpha/H$, if waiting times
are heavy tailed, $\alpha<1$:
\begin{equation}
q(z\approx0)\sim\left\{\begin{array}{ll}
\mbox{const}\cdot|z|^{-1+\alpha/H},&\alpha/H<1,\\
\mbox{const}\cdot\log|z/2|,&\alpha/H=1,\\
q(0)-\mbox{const}\cdot|z|^{-1+\alpha/H},&1<\alpha/H<3,\\
q(0)-\mbox{const}\cdot z^2\log|z/2|,&\alpha/H=3,\\
q(0)-\mbox{const}\cdot z^2,&\alpha/H>3.
\end{array}\right.
\label{pdfsmall}
\end{equation}
The constants depend on the parameters $\mu,\alpha,H$, but not on the scaling variable $z$. Thus, the behaviour around the origin can be divergent ($\alpha/H\leq1$), continuous with divergent derivative ($1<\alpha/H<2$), continuous with discontinuous first derivative ($2\leq\alpha/H\leq3$), and continuous with vanishing
first derivative ($\alpha/H>3$). While the cusp-like shape for low values of
$\alpha/H$ is reminiscent of CTRW propagators, the increasingly smoother shape
for higher values of $\alpha/H$ is imitating Gaussian distributions.
Also note that in the absence of heavy-tailed waiting times, corresponding to 
$\alpha\rightarrow1$, the scaling function returns to the class of stable laws, 
which are completely smooth (i.e., infinitely differentiable) everywhere.
Example plots are given in Fig~\ref{fig.pdf}.

In contrast, if $\mu<2$, we find that heavy tails are directly inherited from
the underlying jump length distribution,
\begin{equation}
q(z\rightarrow\infty)\simeq|z|^{-1-\mu},\quad\mbox{for }\mu<2.
\label{pdflarge}
\end{equation}
This holds regardless of which type of correlations or waiting time
distributions characterize the motion, see also Fig.~\ref{fig.pdf}. In the
special case of Gaussian jump lengths, the tails of the PDF are of
exponential type, $\log[q(z\rightarrow\infty)]\simeq -|z|^{1/2+H(1-\alpha)/\alpha}$.

Finally, let us point out an interesting, but maybe not intuitively expected
property of the scaling function $q(z)$. From Eq.~(\ref{pdffoxH}) one can
derive~\cite{FoxH}
\begin{equation}
q(z)|_{\alpha\rightarrow1}=q(z)|_{H\rightarrow0}.
\label{pdfspecial}
\end{equation}
The limit $\alpha\rightarrow1$ leads back to a steady time progression, 
ultimately rendering internal time and laboratory time equivalent. 
Interestingly, when we study the shape of the PDFs, this is effectively the
same as choosing $H$ very small. Thus, if either waiting times are sufficiently
persistent, or jump distances are sufficiently antipersistent, then the shape
of the propagator indicates dynamics devoid of any stalling or trapping
mechanisms.

\section{Comparison with other models of correlated motions}\label{comparison}

We now briefly compare, contrast and connect the correlated continuous time random walk model discussed in the previous sections to other existing models of correlated motion.

First, we stress that CCTRWs are distinct from the correlated (persistent) random walk models as discussed in~\cite{Turchin1998,Kareiva1983,Bovet1988,Bartumeus2008}. The latter are two-dimensional random walk models, aiming at describing animal foraging and movements patterns. Angular correlations are introduced by means of nonuniform angular distributions, governing the directional evolution of the random walk at each step. Angular and step length distribution define characteristic correlation scales, beyond which the dynamics are essentially Brownian.

The present CCTRW model is a direct continuation of the CTRW with correlated waiting times presented in~\cite{Chechkin2009}. The authors discuss a laboratory time process (see Eqs.~(23) and~(24) in~\cite{Chechkin2009}, we slightly adopt the notation to our needs)
\begin{equation}
 T(s) = \int_0^s m(s-s') \,\mathrm{d}L^+_\alpha(s') ,\qquad\mbox{with}\qquad m(s)=\int_0^{s} M(s') \,\mathrm{d}s' .\nonumber\\
\end{equation}
While the the correlation kernel $m(s)$ defines the integral representation of laboratory time $T(s)$, the function $M(s)=\mathrm{d}m/\mathrm{d}s$ can be interpreted as a correlation kernel for the noise or waiting time process `` $\mathrm{d}T/\mathrm{d}s$'' (see Eq.~(20) of~\cite{Chechkin2009}). Two different types of correlation kernels are taken into consideration. First, power law correlated waiting times, $M(s)\propto s^{-\beta}$, $\beta<1$, lead to a power law correlated laboratory time process, $m(s)\propto s^{1-\beta}$. By identifying $G=1-\beta+1/\alpha$, $G>1/\alpha$, we exactly arrive at the process definition used here, Eq.~(\ref{T}). Since jump lengths in the model of~\cite{Chechkin2009} are Gaussian and independent (which, in our language, means $\mu=2$, $K=1/2$) we expect a scaling relation $X(t)\sim t^H=t^{K/G}=t^{\alpha/[2\alpha(1-\beta)+2]}$. This is fully consistent with the mean squared displacement analysis in Eq.~(43) of~\cite{Chechkin2009}. A second interesting choice for the kernel behaviours is an exponentially decaying one, i.e. $M(s)\propto \exp(-\Delta s)$, $\Delta>0$, corresponding to $m(s)\propto 1- \exp(-\Delta s)$. While the full scaling behaviour is difficult to calculate explicitely, we can look at the limiting cases $t\ll\Delta$ and $t\gg\Delta$. By virtue of the monotonic increase of the process $T(s)$, this is equivalent to studying approximations with respect to internal time $s$. For small $s$, we have $m(s)\simeq s= s^{1-0}$, while for large $s$, we get $m(s)\simeq1=s^{1-1}$. Hence, we expect a turnover from the scaling $X(t)\sim t^{\alpha/[2\alpha+2]}$ at $t\ll\Delta$ to $X(t)\sim t^{\alpha/2}$ at $t\gg\Delta$. This is in perfect agreement with the mean squared displacement results Eqs.~(36) and~(38) in~\cite{Chechkin2009}.

Finally, we wish to draw the connection to the correlated CTRW introduced in Ref.~\cite{Tejedor2010}. On the discrete random walk level, the basic idea is to define a nonstationary and correlated sequence of jump lengths or waiting times in terms of separate random walk processes. For instance, correlated jump lengths $\delta y_n$ are derived from a L\'evy flight in jump length space. In other words, 
\begin{eqnarray}
 \delta y_n &=& \sum_{j=1}^n \xi_j ,\nonumber\\
 Y_n &=& \sum_{j=1}^n \delta y_j = \sum_{j=1}^n\sum_{k=1}^j \xi_k ,
 \label{corrjumps_discrete}
\end{eqnarray}
where the $\xi_j$ are independent, symmetric $\mu$-stable random variables. The intuitive way of guessing a long time limit approximation of this process can be found by replacing sums with integrals: 
\begin{equation}
 Y(s)=\int_0^s \int_0^{s'} \mathrm{d}L_\mu(s'') \,\mathrm{d}s' = \int_0^s L_\mu(s') \,\mathrm{d}s'.
 \label{corrjumps_continuous}
\end{equation}
Indeed the convergence in distribution of the discrete random walk $Y_n$ to the continuous process $Y(s)$ was proved in~\cite{Magdziarz2012}. The latter can be thought of, according to above equation, as an integrated symmetric L\'evy flight. Now according to the spectral analysis discussion we brought up in section~\ref{stat}, such process should actually be included in the class of correlated motions discussed in the present paper. Indeed, we could also rewrite the double sum in Eq.~(\ref{corrjumps_discrete}) as
\begin{equation}
 Y_n = \sum_{j=1}^n (n-j) \,\xi_j .
\end{equation}
The analogous step for continuous time is formal integration by parts of Eq.~(\ref{corrjumps_continuous}):
\begin{equation}
 Y(s)=\int_0^s (s-s') \,\mathrm{d}L_\mu(s').
\end{equation}
Up to a constant prefactor, this exactly corresponds to our definition~(\ref{Y}) with $K=1/\mu +1$. We thus find, in complete accordance with~\cite{Tejedor2010,Magdziarz2012}, that the integrated L\'evy flight is a $\mu$-stable process with superdiffusive scaling $Y(s)\sim s^K=s^{1/\mu+1}$.

Correlated waiting times however, are defined in~\cite{Tejedor2010,Magdziarz2012} in a slightly different manner. There, consecutive waiting times $\delta t_n$ are taken from a symmetric L\'evy flight, subject to a reflecting boundary condition at $\delta t_n=0$. In short,
\begin{eqnarray}
 \delta t_n &=& \left|\sum_{j=1}^n \zeta_j\right| ,\nonumber\\
 T_n &=& \sum_{j=1}^n \delta t_j = \sum_{j=1}^n \left|\sum_{k=1}^j \zeta_k\right| ,
 \label{corrwaits_discrete}
\end{eqnarray}
where the $\zeta_j$ are independent, symmetric $\alpha$-stable random variables with $0<\alpha\leq2$. The continuous version is
\begin{equation}
 T(s)=\int_0^s \left|\int_0^{s'} \mathrm{d}L_\alpha(s'')\right| \,\mathrm{d}s' = \int_0^s \left|L_\alpha(s')\right| \,\mathrm{d}s',
 \label{corrwaits_continuous}
\end{equation}
which is \emph{not} a stable process~\cite{Magdziarz2012}, and hence cannot be represented by any of our correlated laboratory time processes~(\ref{T}). Still, there is a formal analogy in scaling behaviours. It is easy to show that the integrated L\'evy flight on the positive half-line, Eq.~(\ref{corrwaits_continuous}), is self-similar with $T(s)\sim s^{1/\alpha+1}$. The present model yields the same scaling for $G=1/\alpha+1$; interestingly, this corresponds to a single integration of a one-sided $\alpha$-stable motion. In the case of independent Gaussian jump lengths, $\mu=2$ and $K=1/2$, such scaling produces subdiffusive dynamics $X(t)\sim t^{K/G}=t^{\alpha/[2(1+\alpha)]}$, as previously found in~\cite{Tejedor2010}.

\section{Conclusions}\label{conclusion}

The correlated CTRW we introduced here combines the effects of displacements
with infinite variance, sojourn times with infinite mean and long-range
temporal correlations. It is thus applicable to a wide range of complex,
heterogeneous systems. We found that the probability density function is very
distinct from an ordinary Gaussian distribution. We studied its shape extensively, revealing information contained in the tail properties and the detailed behaviour around the origin. However, care must be taken when assessing the effects of correlations: processes with contrasting jump length and waiting time correlations can be indiscernible on the level of scaling and propagator analysis.

Moreover, we classified correlated CTRWs in the context of processes with stationary increments of higher order. Such considerations indicate an intimate connection between strong correlations and higher-, possibly fractional-order integrals of stochastic noise processes.

Further studies of this process should include an in-depth discussion on the 
actual correlations within the correlated model. This question is particularly 
intricate for scale-free displacements ($\mu<2$), since the ordinary correlation
function $\langle X(t_1)X(t_2)\rangle$ is ill-defined. Moreover, it would be
interesting to study a process where correlations within displacements and
waiting times are coupled, such as L\'evy walks. Finally, the physical properties
of such processes such as aging or (non-)ergodicity will be of interest.

\appendix

\section{Asymptotic distributional stationarity of higher order increments}\label{app:stationarity}
We establish here the asymptotic behaviour of increments of arbitrary order for the correlated stable motion $Y(s)$ as defined through Eq.~(\ref{Y}). Recall that by the term ASD we designate the asymptotic long-time stationarity of a single-time distribution.

First, rewrite the correlated motion $Y(s)$, Eq.~(\ref{Y}), as
\begin{equation}
 Y(s)=(\mu K)^{1/\mu}\int_0^\infty M(s-s') \,\mathrm{d}L_\mu(s'), \qquad M(s)=\theta(s) \,s^{K-1/\mu} ,
 \label{app:kernel}
\end{equation}
with $\theta$ denoting the Heaviside step function, i.e. $\theta(s\geq0)=1$ and  $\theta(s<0)=0$.

The $n$-th order increments, Eq.~(\ref{Y_nincrements}), have a similar stochastic integral representation, namely
\begin{equation}
 \Delta^{(n)} Y(s;\tau_1,...,\tau_n) = (\mu K)^{1/\mu}\int_0^\infty M^{(n)}(s-s') \,\mathrm{d}L_\mu(s')
\end{equation}
with associated integration kernels
\begin{eqnarray}
  M^{(1)}(s;\tau) &=& M(s+\tau) - M(s) \nonumber\\
  M^{(2)}(s;\tau_1,\tau_2) &=& M^{(1)}(s+\tau_2;\tau_1) - M^{(1)}(s;\tau_1) \nonumber\\
  &&\vdots\nonumber\\
  M^{(n)}(s;\tau_1,...,\tau_n) &=& M^{(n-1)}(s+\tau_n;\tau_1,...,\tau_{n-1}) - M^{(n-1)}(s;\tau_1,...,\tau_{n-1}) .\nonumber\\
  \label{app:incrementkernel}
\end{eqnarray}
The characteristic function of the distribution of $n$-th order increments is related through
\begin{eqnarray}
 &&\log\left\{ \langle\exp\left[ik\Delta^{(n)} Y(s;\tau_1,...,\tau_n)\right]\rangle \right\} \nonumber\\
 &&\quad= -\mu K|k|^\mu\int_0^\infty \left|M^{(n)}(s-s';\tau_1,...,\tau_n)\right|^\mu \,\mathrm{d}s' \nonumber\\
 &&\quad= -|k|^\mu\left(I^{(n)}_{\mu,K}(s)-I^{(n)}_{\mu,K}(-\infty)\right)
\end{eqnarray}
with
\begin{equation}
 I^{(n)}_{\mu,K}(s)= \mu K\int_0^s \left|M^{(n)}(s';\tau_1,...,\tau_n)\right|^\mu \,\mathrm{d}s' .
 \label{app:incrementpdf} 
\end{equation}

The question of whether or not such distribution has a nontrivial limit for $s\rightarrow\infty$ is determined by the integral $I^{(n)}_{\mu,K}(s)$ and hence by the tail asymptotics of the integration kernel $M^{(n)}(s)$. Notice that the step function $\theta(s)$ in the process kernel~(\ref{app:kernel}) passes on to the increment kernels~(\ref{app:incrementkernel}) and contributes in the form $\theta(s+\tau_1)$, $\theta(s+\tau_2)$,..., $\theta(s+\tau_1+\tau_2)$, $\theta(s+\tau_1+\tau_3)$,..., \textit{etc}. It thus defines several \emph{lower} bounds for the integral $I^{(n)}_{\mu,K}(-\infty)$. Conversely, for $s\geq0$, all step functions entering the integral $I^{(n)}_{\mu,K}(s)$ are identically equal unity. At this point, we have to distinguish two parameter classes:

First, we can have $K=1/\mu+m$ for some nonnegative integer $m$. Then the process kernel is $M(s\geq0)= s^m$, and we can use standard polynomial calculus. One can show that for all $n\leq m$ increment kernels $M^{(n)}$ are polynomials of degree $m-n$ and thus the nonstationary contribution $I^{(n)}_{\mu,K}(s)$ grows indefinitely for large $s$. Conversely, for $n\geq m+1$, $M^{(n)}$ vanishes identically and so does the nonstationary contribution to the characteristic function. In particular, we have the L\'evy stable motions, $m=0$, $K=1/\mu$, with stationary increments of all orders $n\geq1$.

Now consider the second class of parameter pairs, i.e. $K\neq1/\mu+m$ for all nonnegative integers $m$. In this case, the process kernel is a noninteger power law, and we recursively find tail asymptotics
\begin{eqnarray}
  M^{(n)}(s;\tau_1,...,\tau_n) \sim \tau_n \, \left(\frac{\partial M^{(n-1)}}{\partial s}\right)(s;\tau_1,...,\tau_{n-1})
\end{eqnarray}
for $s\gg\tau_1+...+\tau_n$. This implies the explicit tail behaviour,
\begin{eqnarray}
 M^{(n)}(s;\tau_1,...,\tau_n) &\sim& \tau_1...\tau_n \, \left(\frac{\partial^n M}{\partial s^n}\right)(s) \nonumber\\
 &=& \frac{\Gamma(K-1/\mu)}{\Gamma(K-1/\mu-n)} \, \tau_1...\tau_n \, s^{K-1/\mu-n} ,
\end{eqnarray}
so for the ultimate time dependence of the nonstationary part of the characteristic function, we get
\begin{eqnarray}
 I^{(n)}_{\mu,K}(s) &\sim& \mu K\left|
 \frac{\Gamma(K-1/\mu)}{\Gamma(K-1/\mu-n)} \, \tau_1...\tau_n \right|^\mu \nonumber\\
 &&\quad\times \left\{\begin{array}{ll}
  \mbox{const} ,&\mbox{ for } n>K ,\\
  \log(s) ,&\mbox{ for } n=K ,\\
  \left[\mu (K-n)\right]^{-1}\,s^{\mu (K-n)} ,&\mbox{ for } n<K.
 \end{array}\right.
\end{eqnarray}
Hence, only increments of order $n>K$ are ASD, while all lower order increments are spreading indefinitely with time.

\ack

We aknowledge funding from the Academy of Finland (FiDiPro scheme) and the
CompInt graduate school of TU Munich.

\end{document}